# Spatially Resolving Spin-split Edge States of Chiral Graphene Nanoribbons

Chenggang Tao[1, 2*], Liying Jiao[3*], Oleg V. Yazyev[1, 2*], Yen-Chia Chen[1, 2], Juanjuan Feng[1], Xiaowei Zhang[1, 2], Rodrigo B. Capaz[1, 5], James M. Tour[4], Alex Zettl[1, 2], Steven G. Louie[1, 2†], Hongjie Dai[3†], Michael F. Crommie[1, 2†]

[1] *Department of Physics, University of California at Berkeley, Berkeley, CA 94720, USA*

[2] *Materials Science Division and Chemical Science Division, Lawrence Berkeley National Laboratory, Berkeley, CA 94720, USA*

[3] *Department of Chemistry and Laboratory for Advanced Materials, Stanford University, Stanford, CA 94305, USA*

[4] *Department of Chemistry, and Smalley Institute for Nanoscale Science and Technology, Rice University, TX 77005, USA*

[5] *Instituto de Física, Universidade Federal do Rio de Janeiro, Caixa Postal 68528, Rio de Janeiro, RJ 21941-972, Brazil*

\* *These authors contributed equally to this work*

† *Correspondence and requests for materials should be addressed to these authors*

A central question in the field of graphene-related research is how graphene behaves when it is patterned at the nanometer scale with different edge geometries. Perhaps the most fundamental shape relevant to this question is the graphene nanoribbon (GNR), a narrow strip of graphene that can have different chirality depending on the angle at which it is cut. Such GNRs have been predicted to exhibit a wide range of behaviour (depending on their chirality and width) that includes tunable energy gaps [1] and the presence of unique one-dimensional (1D) edge states with unusual magnetic structure [2-7]. Most GNRs explored experimentally up to now have been characterized via electrical conductivity, leaving the critical relationship between electronic structure and local atomic geometry unclear (especially at edges) [8,9]. Here we present a sub-nm-resolved scanning tunnelling microscopy (STM) and spectroscopy (STS) study of GNRs that allows us to examine how GNR electronic structure depends on the chirality of atomically well-defined GNR edges. The GNRs used here were chemically synthesized via carbon nanotube (CNT) unzipping methods that allow flexible variation of GNR width, length, chirality, and substrate [10,11]. Our STS measurements reveal the presence of 1D GNR edge states whose spatial characteristics closely match theoretical expectations for GNR's of similar width and chirality. We observe width-dependent splitting in the GNR edge state energy bands, providing compelling evidence of their magnetic nature. These results confirm the novel electronic behaviour predicted for GNRs with atomically clean edges, and thus open the door to a whole new area of applications exploiting the unique magnetoelectronic properties of chiral GNRs [6].

The chirality of a GNR is characterized by a chiral vector (n, m) or, equivalently, by chiral angle $\theta$, as schematically shown in Fig. 1a. (*n*, *m*) expresses the GNR edge orientation in graphene lattice coordinates while $\theta$ is the angle between the zigzag direction and the actual edge orientation. GNRs having different widths and chiralities were deposited onto a clean Au(111) surface and measured using STM. Fig. 1b shows a room temperature image of a single monolayer GNR (GNR height is determined from linescans such as that shown in Fig. 1b inset; some multilayer GNRs were observed, but we focus here on monolayer GNRs). The GNR of Fig. 1b has a width of 23.1 nm, a length greater than 600 nm, and exhibits straight, atomically smooth edges (the highest quality GNR edges, such as those shown in Figs.1 and 2, were observed in GNRs synthesized as in ref. [10]). Such GNRs are seen to have a "bright stripe" running along each edge.

This stripe marks a region of curvature near the terminal edge of the GNR which has a maximum extension of 2 to 3 Å above the mid-plane terrace of the GNR, and a width of ~ 30 Å (see line scan in Fig. 1b inset). Such edge-curvature was observed for all high quality GNRs examined in this study (more than 150, including GNRs deposited onto a Ru(0001) surface). This is reminiscent of curved edge structures observed previously near graphite step-edges [12]. We rule out that these GNRs are crushed nanotubes both from the GNR height and from the measured ratio (observed to be π) for GNR width to nanotube height for partially unzipped CNTs (see Supplementary Information). Low temperature STM images (Figs. 1c, 2a) reveal finer structure in both the interior GNR terrace and the edge region. Fig. 2a, for example, shows the atomically-resolved edge region of a monolayer GNR and clearly exhibits how the periodic graphene sheet of the GNR terminates cleanly and with atomic order at the gold surface.

Such high-resolution images allow us to experimentally determine the chirality of GNRs, and to create structural models of observed edge regions. In Fig. 2a, for example, we see rows of protrusions (with a spacing of ~ 2.5 Å) near the edge of a GNR having width = 19.5 ± 0.4 nm. These protrusions have the spacing expected for adjacent graphene hexgons, and thus the orientation of the observed rows determines the zigzag direction. By comparing this row orientation with the GNR outer edge orientation we are able to extract the GNR chirality (details in Supplementary Information). The GNR displayed in Fig. 2a has an (8, 1) chirality (equivalent to $\theta = 5.8°$), and the resulting structural model for this GNR is shown in Fig. 2b. We find the distribution of GNR chiralities to be essentially random. This is consistent with our structural data which indicates that the CNT unzipping direction is very close to the axial direction of the precursor CNTs (see Supplementary Information), as well as the fact that the precursor CNTs have a broad chirality distribution [13].

We explored the local electronic structure of GNR edges using STS, in which $dI/dV$ measurement reflects the energy-resolved local density of states (LDOS) of a GNR. Figs. 2c and 2d show $dI/dV$ spectra obtained at different positions (as marked) near the edge of the (8, 1) GNR pictured in Fig. 2a. $dI/dV$ spectra measured within 24 Å of the GNR edge typically show a broad gap-like feature having an energy width of ~130 meV. This is very similar to behaviour observed in the middle of large-scale graphene sheets, and is attributed to the onset of phonon-assisted inelastic electron tunnelling [14] for $|E| \geq$ 65 meV. This feature disappears further into the interior of the GNR, as expected due to increased tunnelling to the Au substrate [15]. Very close to the GNR edge, however, we observe additional features in the spectra. The most dominant of these features are two peaks that rise up within the elastic tunnelling region (i.e. at energies below the phonon-assisted inelastic onset) and which straddle zero bias. For the GNR shown in Fig. 2a (which has a width of 19.5 ± 0.4 nm) the two peaks are separated in energy by a splitting of $\Delta = 23.8 \pm 3.2$ meV. Similar energy-split edge state peaks have been observed in all clean chiral GNRs that we investigated spectroscopically at low temperature. For example, the inset to Fig. 2c shows a higher resolution spectrum exhibiting energy-split edge state peaks for a (5, 2) GNR having a width of 15.6 nm and an energy splitting of $\Delta = 27.6 \pm 1.0$ meV. The two edge state peaks are often asymmetric in intensity (depending on specific location within the GNR edge region), and their mid-point is often slightly offset from $V_s = 0$ (within a range of ± 20 meV). As seen in the spectra of Fig. 2c, the amplitude of the peaks grows as one moves closer to the terminal edge of the GNR, before falling abruptly to zero as the carbon/gold terminus is crossed into the gold surface. The spatial dependence of the edge state peak amplitude as one moves perpendicular from the GNR edge is plotted in Fig. 3a and shows exponential behaviour. The edge state spectra also vary as one moves parallel to the GNR edge, as shown in Fig. 2d. The parallel dependence of the edge state peak amplitude is plotted in Fig. 3b, and is seen to oscillate with an approximate 20 Å period, corresponding closely to the 21 Å periodicity of an (8, 1) edge.

    We have also characterized monolayer GNRs having different chiralities and widths. In Fig. 3c, we plot the width dependence of the measured energy gap of GNR

edge states for a broad range of chirality (3.7° < $\theta$ < 16.1°). The measured edge state energy splitting shows a clear inverse correlation with GNR width.

The high quality of the atomically well-defined edge structures observed here allows us to quantitatively compare our experimental data to theoretical calculations of the electronic structure of chiral GNRs. We find that the spectroscopic features we observe correspond closely to the spatial and energy-dependence predicted for 1D magnetic edge states coupled across the width of a chiral GNR. This behaviour is quite different from the properties observed previously for graphite step edges, armchair nanoribbons, and comparatively less ordered graphene platelet edges where no magnetism-induced energy splitting has been seen [16-19].

In order to compare our experimental data with theoretical predictions for GNRs, we used a Hubbard model Hamiltonian solved self-consistently within the mean-field approximation [5] for an (8, 1) GNR having the same width as the actual (8, 1) GNR shown in Fig. 2a. The Hubbard model Hamiltonian:

$$H = -t \sum_{\langle ij \rangle, \sigma} [c_{i\sigma}^\dagger c_{j\sigma} + \text{h.c.}] + U \sum_i n_{i\uparrow} n_{i\downarrow} \qquad (1)$$

consists of a one-orbital nearest-neighbor tight-binding Hamiltonian (first term) with an on-site Coulomb repulsion term (the latter term leads to magnetic ordering). In this expression $c_{i\sigma}^\dagger$ and $c_{j\sigma}$ are operators that create and annihilate an electron with spin $\sigma$ at the nearest neighbor sites $i$ and $j$ respectively, $t = 2.7$ eV is a hopping integral [20,21], $n_{i\sigma} = c_{i\sigma}^\dagger c_{i\sigma}$ is the spin-resolved electron density at site $i$, and $U$ is an on-site Coulomb repulsion. This GNR model is defined only by the π-bonding network. The terminal σ-bonds at the GNR edges are considered to be passivated and do not alter the π-system (this should, in general, correctly model a range of different possible edge-adsorbate bonding configurations [6,22]). The out-of-plane curvature seen experimentally near GNR edges is not included in this model since the measured radii of curvature are sufficiently large (>20 Å) that they are not expected to significantly affect GNR electronic structure [23] (we tested this conjecture by including the observed curvature in some calculations, and found that it has no significant effect - either via σ-π coupling or via pseudofield effects - on the calculated GNR electronic structure). The effect of the gold substrate here is taken only as a charge reservoir that can slightly shift the location of $E_F$ within the GNR band structure and reduce the magnitude of the effective $U$ parameter via electrostatic screening (the experimental charge-induced energy shifts seen here are within the range of charge-induced energy shifts observed previously for CNTs on Au [24]).

We first calculated the GNR electronic structure for $U = 0$, which effectively omits the electron-electron interactions responsible for the onset of magnetic correlations. This results in the theoretical band structure and density of states (DOS) shown in Figs. 4a, b (blue dashed lines). The finite width of the GNR leads to a family of sub-bands in the band structure, with no actual band gap (Fig. 4a). A flat band at $E = 0$ due to localized edge states spans the entire 1D Brillouin zone for the (8, 1) GNR, leading to a strong van Hove singularity (i.e., a peak) in the DOS at $E = 0$ (Fig. 4b). The DOS in this case does not resemble what is seen experimentally. We next calculated the (8, 1) GNR electronic structure for $U > 0$. Here the electron-electron interactions lift the degeneracy of the edge states by causing ferromagnetic correlations to develop along the GNR edges and antiferromagnetic correlations to develop across the GNR. This leads to a spin-polarization of the edge states that splits the single low-energy peak seen in the $U = 0$

DOS into a series of van Hove singularities, thus opening up a gap at $E = 0$. Such behaviour is seen in the band structure and DOS of Figs. 4a, b (solid red lines). We identify the lowest-energy pair of van Hove singularities with the pair of peaks observed experimentally near zero bias for GNR edges. We focus our experiment/theory comparison to the low-energy regime ($|E| \leq 65$ meV) because higher energy experimental features are complicated by the onset of phonon-assisted inelastic tunnelling [25] (the low-energy edge state peaks, by contrast, do not have the characteristics of inelastic modes).

We find that our experimental spectroscopic edge state data for the (8, 1) GNR is in agreement with model Hamiltonian calculations for $U = 0.5t$. The theoretical band gap of 29 meV is very close to the experimentally observed value of $23.8 \pm 3.2$ meV (the value of $U$ used here is lower than a value obtained previously from a first-principles calculation [20,21], presumably due to screening from the gold substrate). Our experimentally observed energy-split spectroscopic peaks thus provide compelling evidence for the formation of spin-polarized edge states in pristine GNRs (such splitting does not arise for the non-magnetic $U = 0$ case described above). We are further able to compare the spatial dependence of the calculated magnetic edge states with the experimentally measured STS results. The dashed line in Fig. 3a shows the theoretical local density of states (LDOS) calculated at the energy of the low-energy edge state peaks as one moves perpendicular away from the GNR edge and into the (8, 1) GNR interior. The predicted exponential decay length of ~12 Å is in reasonable agreement with the experimental data (the data does exhibit a steeper slope, possibly due to substrate-induced screening not included in the model). Variation seen in the calculated LDOS of the magnetic edge state in the direction parallel to the GNR edge also compares favorably with our experimental observations (Fig. 3b). This oscillation in edge state amplitude arises from "breaks" in the zigzag edge structure due to the chiral nature of the (8, 1) GNR edge (see edge structure of Fig. 2b).

We are similarly able to compare the GNR width dependence of our experimentally measured edge state gaps to theoretical calculations. Since the measured GNRs having different widths also have different chiralities (over the range $3.7 \pm 0.3° < \theta < 16.1 \pm 2.2°$), we have calculated the theoretical edge state gap vs. width behaviour over a range of chiralities ($0° < \theta < 15°$). The pink shaded region in Fig. 3c shows the results of this calculation for spin-polarized GNR edge states, and is seen to compare favorably with our experimentally observed width-dependent edge state gap. This provides strong evidence that the edge state gap we observe experimentally is not a local effect, as might occur, say, in response to some unknown molecules bound to the GNR edge, but rather depends on the full GNR electronic structure, including interaction *between* the edges (as expected for spin-polarized edge states).

In conclusion, we provide strong experimental evidence for the existence of magnetic edge states in chiral GNRs with atomically well-defined edges. This behaviour shows that it is possible to create new tunable magnetic and electronic nanostructures by producing chiral GNRs with precisely defined crystallographic orientation and edges, thus creating opportunity for a wide range of new electronic and spintronic applications.

**Acknowledgements**

Research supported by the Office of Naval Research Multidisciplinary University Research Initiative (MURI) award nos. N00014-09-1-1066 and N00014-09-1-1064 (GNR sample preparation and characterization, GNR modeling), by the Director, Office of Science, Office of Basic Energy Sciences, Materials Sciences and Engineering Division, of the U.S. Department of Energy (DOE) under contract no. DE-AC02-05CH11231 (STM instrumentation development and measurements), by the National Science Foundation award no. DMR-0705941 (software development to numerically simulate electron correlation effects), by Intel (development of GNR synthesis techniques), and by MARCO MSD (preliminary GNR screening through use of AFM and transport methods). R.B.C acknowledges financial support from Brazilian agencies CNPq, CAPES, FAPERJ, and INCT.

**Figure Legends**

**Figure 1 | Topography of graphene nanoribbons (GNRs) on Au(111)**. **a,** A schematic drawing of an (8, 1) GNR. The chiral vector (*n*, *m*) connecting crystallographically equivalent sites along the edge defines the GNR edge orientation (black arrow). The blue and red arrows are the projections of the (8, 1) vector onto the basis vectors of the graphene lattice. Zigzag and armchair edges have corresponding chiral angle of $\theta$ = 0° and $\theta$ = 30° respectively, while the (8, 1) edge has an chiral angle of $\theta$ = 5.8°. **b,** Constant-current STM image of a monolayer GNR on Au(111) at room temperature ($V_s$ = 1.5 V, *I* = 100 pA). Inset shows the indicated line profile. **c,** Higher resolution STM image of a GNR at T = 7K ($V_s$ = 0.2 V, *I* = 30 pA, grey-scale height map).

**Figure 2 | Edge states of GNRs. a,** Atomically-resolved topography of the terminal edge of an (8, 1) GNR with measured width of 19.5 ± 0.4 nm ($V_s$ = 0.3 V, *I* = 60 pA, T = 7K). **b,** Structural model of the (8, 1) GNR edge shown in **a**. **c,** *dI/dV* spectra of the GNR edge shown in **a**, measured at different points (black dots, as shown) along a line perpendicular to the GNR edge at T = 7K. Inset shows higher-resolution *dI/dV* spectrum for edge of a (5, 2) GNR with width of 15.6 ± 0.1 nm (initial tunnelling parameters $V_s$ = 0.15 V, *I* = 50 pA; wiggle voltage $V_{rms}$ = 2 mV). The dashed lines are a guide to the eye. **d,** *dI/dV* spectra measured at points (red dots, as shown) along a line parallel to the GNR edge

shown in **a** at T = 7K (initial tunnelling parameters for **c** and **d** are $V_s$ = 0.3 V, $I$ = 50 pA; wiggle voltage $V_{rms}$ = 5 mV).

**Figure 3 | Position and width-dependent edge state properties. a,** Solid blue dots show experimental edge state peak amplitude at points along a line perpendicular to the carbon/gold edge terminus (same positions as shown in Fig. 2**c**). Peak amplitude and energies were determined by fitting Lorentzian curves to the two peaks observed in the measured spectra at each location in Fig. 2**c** over the range -30 mV < $V_s$ < 30 mV. The energy positions of these peaks were found to be 6.7 ± 1.6 mV and -17.2 ± 2.2 mV. The positional dependence of the peak amplitude at 6.7 mV is plotted. Error bars (shown when larger than plotted points) reflect the range of Lorentzian parameters that result in a good fit to the data. Dashed red line shows calculated local density of states (LDOS) at locations spaced perpendicular to the edge terminus for an (8, 1) GNR (see text) at the energy of the DOS peak nearest the band-edge. Theoretical LDOS values include a single global constant offset to model the added contribution from Au surface LDOS, and a single global constant multiplicative factor to model the unknown total area of the STM tunnel junction. **b,** Solid blue dots show experimental edge state peak amplitude at locations spaced along a line parallel to the carbon/gold edge terminus (same positions as shown in Fig. 2**d**). The peak amplitude shown here is determined as in **a** for the edge state peak at -17.2 mV. Dashed red line shows theoretical edge state LDOS for an (8, 1) GNR at points parallel to the edge terminus (calculated as in **a**). The edge state LDOS

amplitude oscillates parallel to the edge with a 21 Å period. **c,** Width dependence of the edge state energy gap of chiral GNRs. From left to right, the chiralities of experimentally measured GNRs are (13, 1), (3, 1), (4, 1), (5, 2), and (8, 1) respectively, corresponding to a range of chiral angle 3.7° < $\theta$ < 16.1°. The pink shaded area shows the predicted range of edge state band gaps as a function of width evaluated for chiral angles in the range 0° < $\theta$ < 15° ($U$ = 0.5$t$, $t$ = 2.7 eV) [26].

**Figure 4 | Theoretical band structure and density of states (DOS) of a 20-nm-wide (8, 1) GNR**. **a,** Dashed blue line shows the calculated GNR electronic structure in the absence of electron-electron interactions ($U$ = 0). Solid red line shows the calculated GNR electronic structure for $U$ = 0.5$t$ ($t$ = 2.7 eV). Finite $U$ > 0 splits degenerate edge states at E = 0 into spin-polarized bands opening a band gap (arrows). **b,** Dashed blue line shows the (8, 1) GNR DOS for the $U$ = 0 case in **a**. The peak at $E$ = 0 is due to the degeneracy of edge states in the absence of electron-electron interactions. Solid red line shows the (8, 1) GNR DOS for $U$ = 0.5$t$. The opening of the band gap (arrows) reflects the predicted energy splitting due to the onset of magnetism in spin-polarized edge states for $U$ > 0, and compares favorably with the experimental data for the (8,1) GNR of Fig. 2.

**Figures**

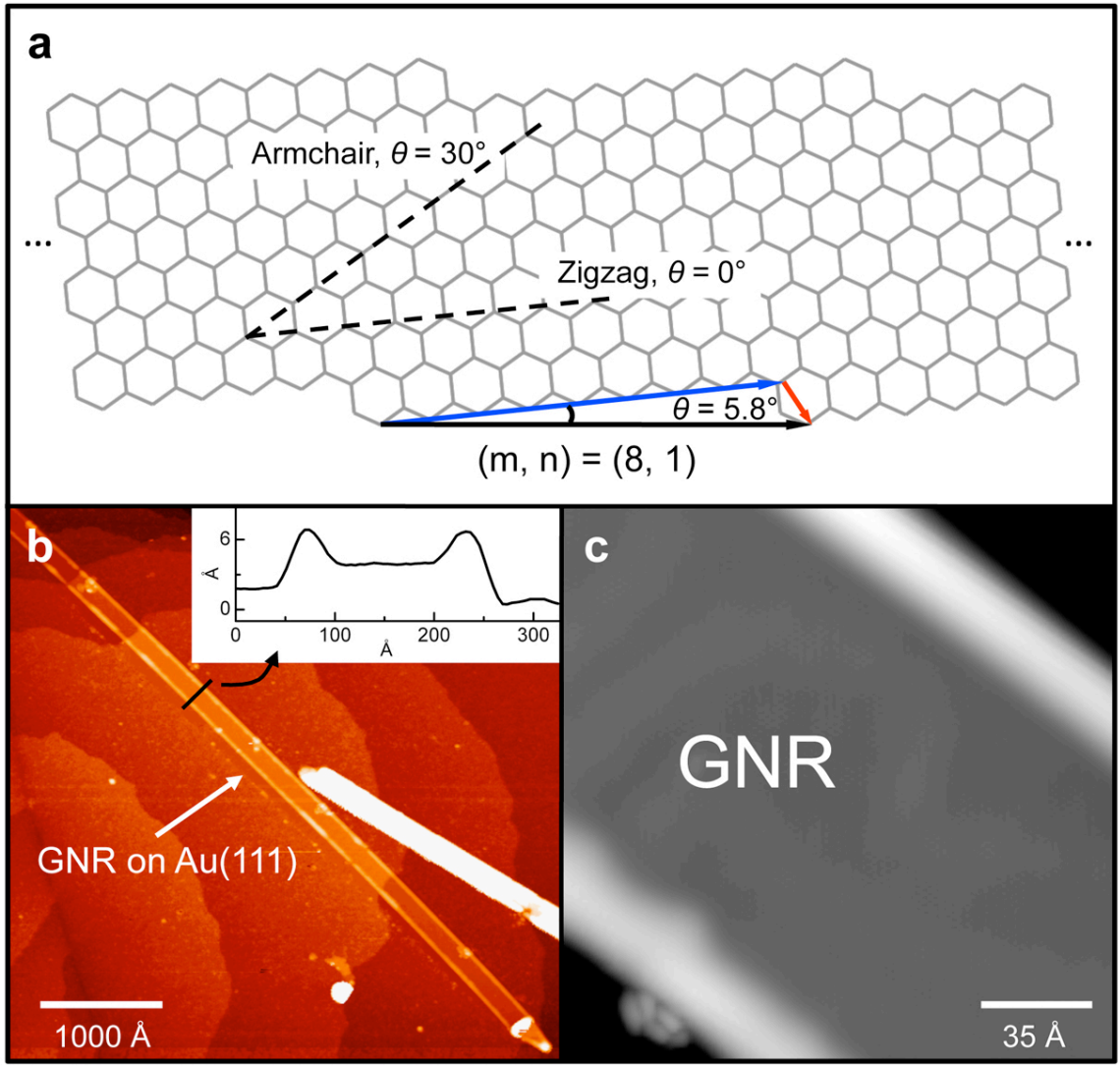

**Figure 1**

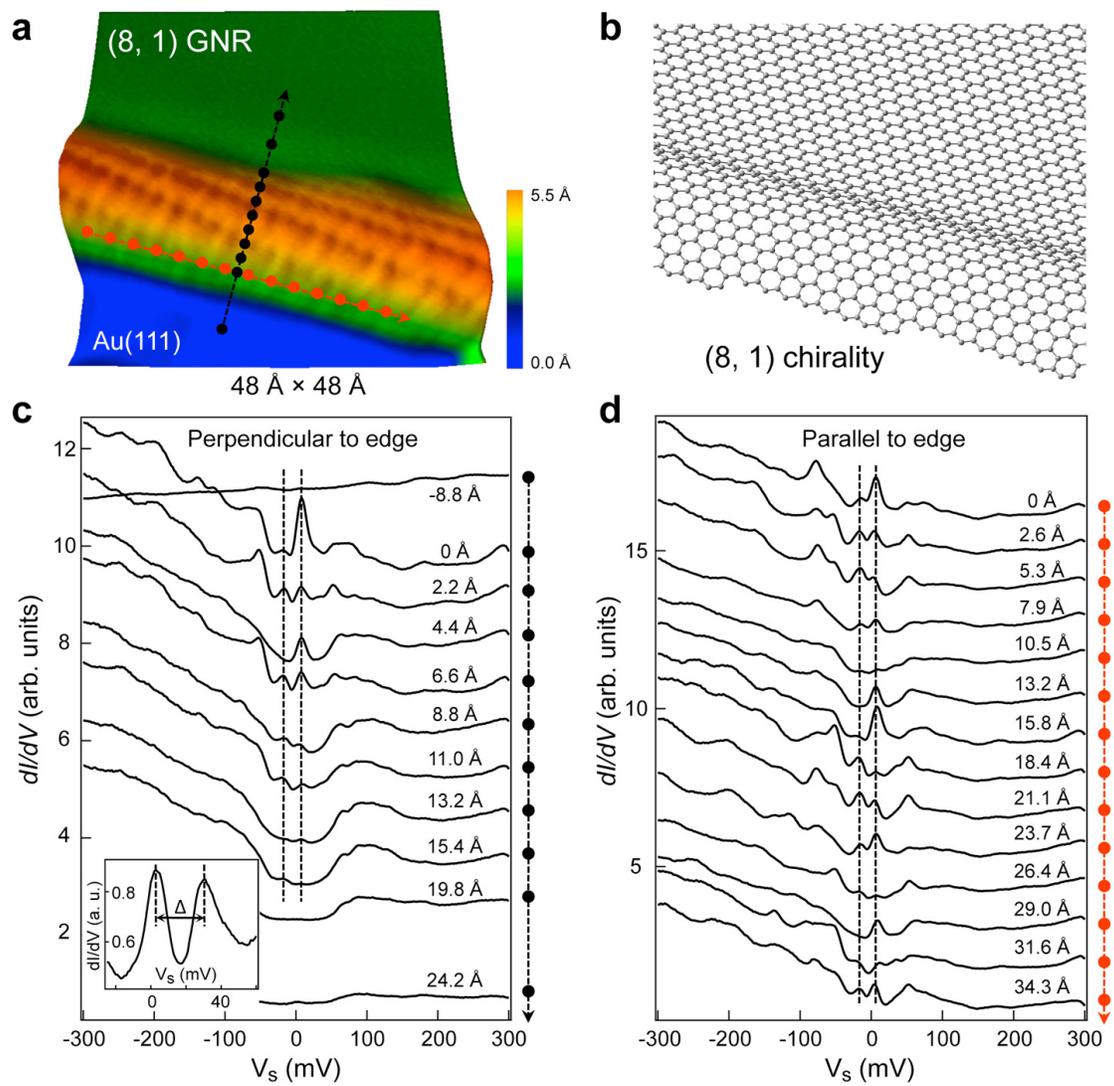

**Figure 2**

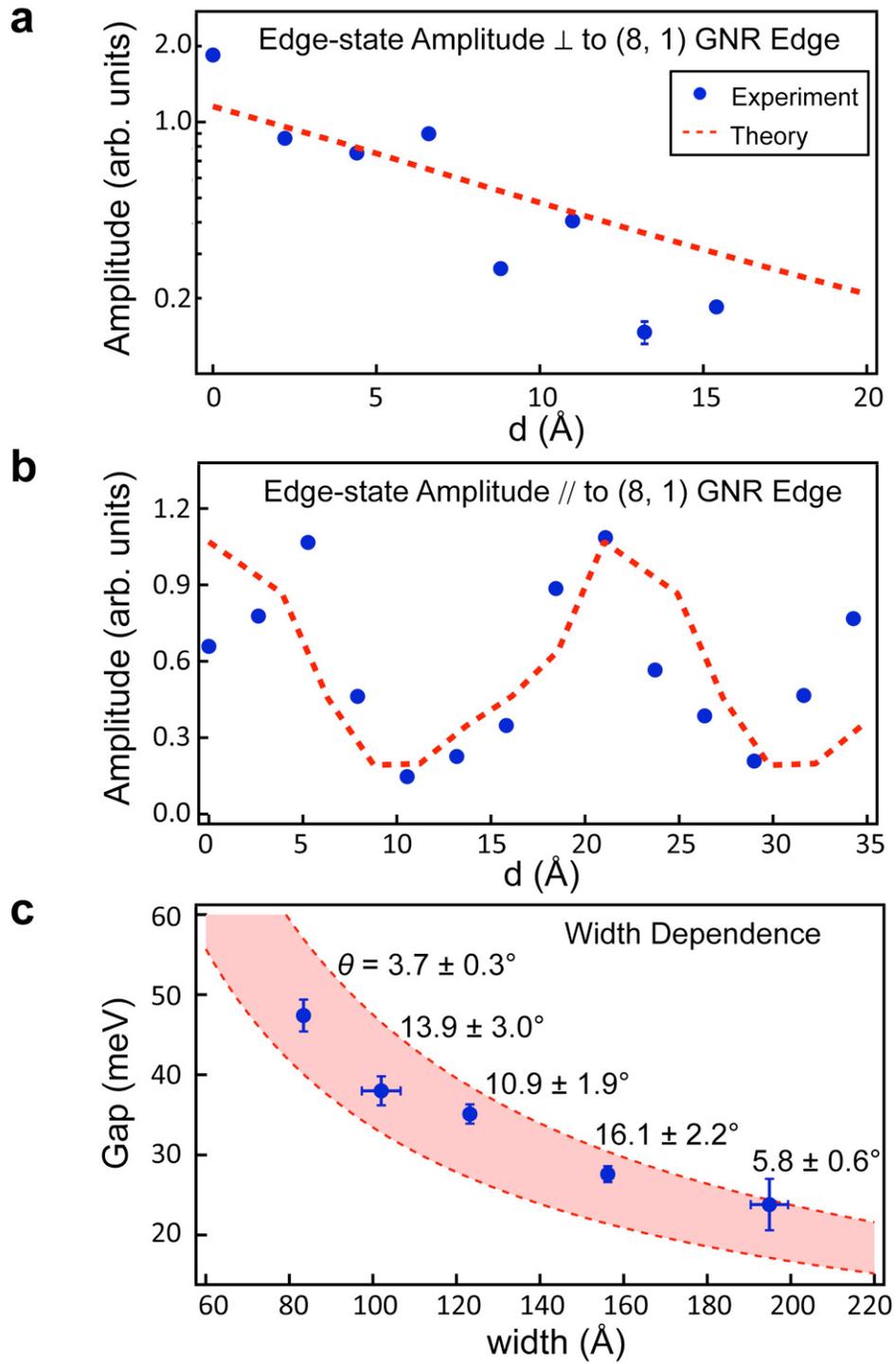

**Figure 3**

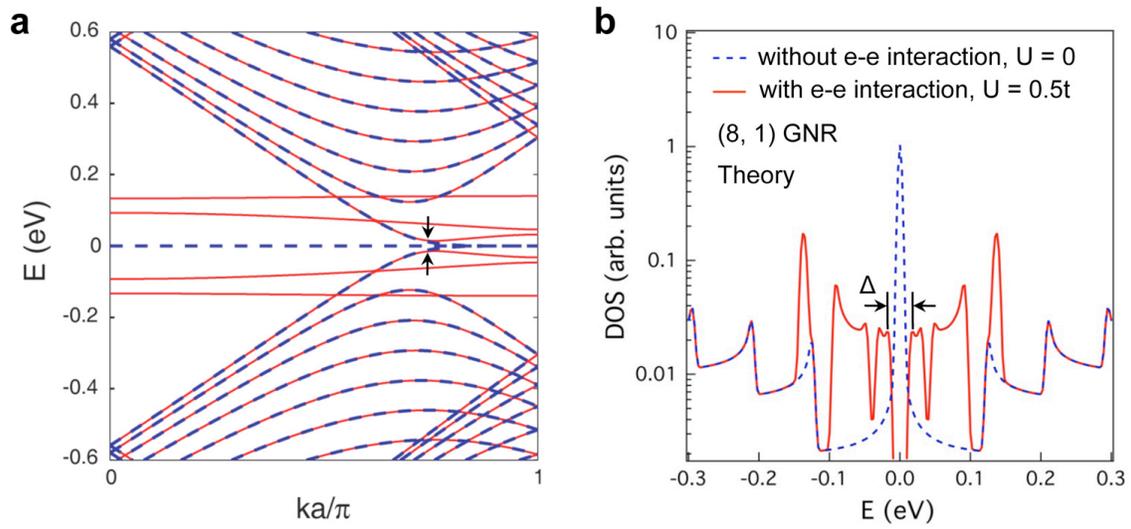

**Figure 4**

# Supplementary Information for "Spatially Resolving Spin-split Edge States of Chiral Graphene Nanoribbons"

Chenggang Tao[1,2*], Liying Jiao[3*], Oleg V. Yazyev[1,2*], Yen-Chia Chen[1,2], Juanjuan Feng[1], Xiaowei Zhang[1,2], Rodrigo B. Capaz[1,5], James M. Tour[4], Alex Zettl[1,2], Steven G. Louie[1,2†], Hongjie Dai[3†], Michael F. Crommie[1,2†]

[1] Department of Physics, University of California at Berkeley, Berkeley, CA 94720, USA

[2] Materials Science Division and Chemical Science Division, Lawrence Berkeley National Laboratory, Berkeley, CA 94720, USA

[3] Department of Chemistry and Laboratory for Advanced Materials, Stanford University, Stanford, CA 94305, USA

[4] Department of Chemistry, and Smalley Institute for Nanoscale Science and Technology, Rice University, TX 77005, USA

[5] Instituto de Física, Universidade Federal do Rio de Janeiro, Caixa Postal 68528, Rio de Janeiro, RJ 21941-972, Brazil

[*] These authors contributed equally to this work

[†] Correspondence and requests for materials should be addressed to these authors

**Contents:**

**1. Materials and sample preparation**

**2. Details of STM and STS measurements**

**3. Characterization of partially unzipped CNTs**

**4. Determination of chirality**

**5. *dI/dV* spectra of a (5, 2) GNR**

**1. Materials and sample preparation**

The GNRs shown in this study were produced by unzipping carbon nanotubes (*S1*). GNRs were deposited onto clean Au(111) surfaces using a spin-coating method. Au(111) substrates were first cleaned by sputtering and annealing in ultra-high vacuum (UHV) before spin-coating. The samples were then transferred into the UHV chamber of our STM system (base pressure ~ $2.0 \times 10^{-10}$ torr). After heat treatment up to 500 °C in UHV, the samples were directly transferred onto the STM stage in the same chamber for measurements.

**2. Details of STM and STS measurements**

STM measurements were performed using a home-built STM held at low temperature (T = 7 K) for maximum spatial and spectroscopic resolution. STM topography was obtained in constant-current mode using a PtIr tip, and *dI/dV* spectra were measured through lock-in detection of the a.c. tunnelling current driven by a 451 Hz, 1 - 5 mV (r.m.s.) signal added to the junction bias (the sample potential referenced to the tip) under open-loop conditions. We also performed large-scale topographic surveys of GNR samples before low temperature measurement. This was done in an Omicron variable temperature STM in UHV at room temperature. The three large-scale topographic STM images shown (Fig. 1b and Figs. S1a-b) were obtained in the Omicron STM at room temperature.

## 3. Characterization of partially unzipped CNTs

During our large-scale topographic survey on GNR samples at room temperature, we observed some partially unzipped CNTs, as shown in Figs. S1a-b. An interesting question for the unzipping process is whether the unzipping direction is along the axis direction or along a spiral direction of the precursor CNTs. From a simple geometric relationship, we know that the ratio between the width of a GNR and the diameter of a CNT for partially unzipped CNTs should be $\pi$ if the unzipping direction is along the axis direction (a schematic model is drawn in Fig. S1c), while the ratio should be smaller than $\pi$ if the unzipping direction is along a spiral direction. We measured the height of the CNT part (denoted as $h$) and the width of the GNR part (denoted as $w$) of the partially unzipped CNTs. The average ratio between the width of the GNR part and the measured height of the CNT part ($w/h$) is $3.2 \pm 0.1$. This average ratio is from 4 partially unzipped CNTs (and on each one, $w/h$ is averaged from more than 10 line profiles of either part). This result implies that the unzipping process is along the axis of the precursor CNTs.

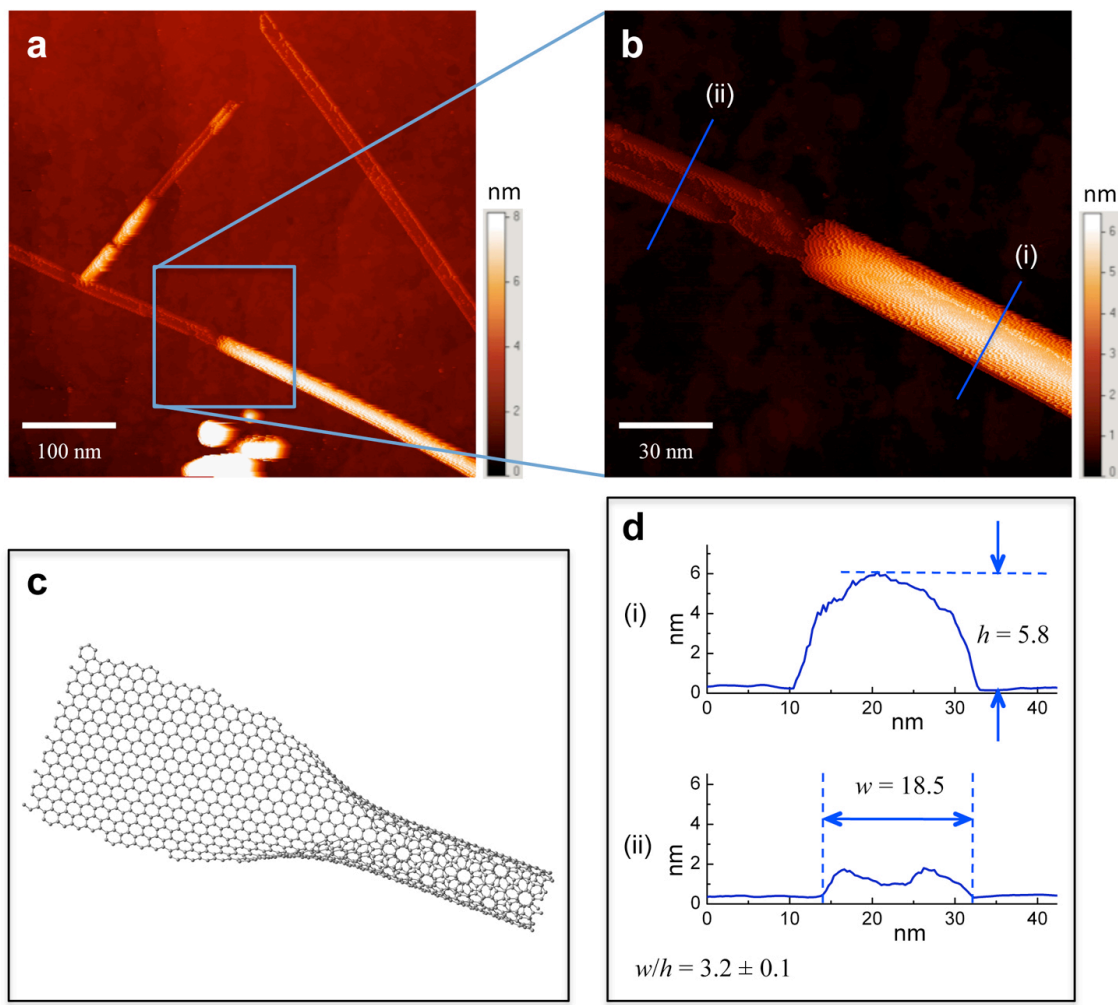

**Figure S1 | Characterization of partially unzipped CNTs. a,** STM image of partially unzipped CNTs ($V_s$ = 1.5 V, I = 100 pA). **b,** A zoom-in image of the GNR-CNT transition region of a partially unzipped CNT ($V_s$ = –1.5 V, I = 100 pA). **c,** Schematic drawing of a partially unzipped CNT with the unzipping direction along the axis direction of a precursor CNT. **d,** Line profiles of the GNR part and the CNT part of the partially unzipped CNT in **b**. The corresponding positions of the line profiles are marked by blue lines in **b**.

## 4. Determination of chirality

Fig. S2 shows how we determine the chirality of GNRs in our measurement. In this STM image, the lower part is the gold surface and the upper part is the GNR. The color contrast is optimized for showing periodic structure near the edge. We can clearly see rows of protrusions in the direction of the yellow line. In each row the protrusions have a spacing ~ 2.5 Å, indicating the rows are along the zigzag direction. Another zigzag direction is along the blue line. The terminal edge orientation can be precisely determined since the GNRs have atomically smooth edges, as shown here and in the large-scale images as well (Fig. 1b and c). Here the black line indicates the terminal edge orientation. The two ends of the back line connect adjacent equivalent sites, which can be determined by the relative position of the black line on the dim dots. By simply projecting this edge vector onto the basis vectors of graphene lattices (the yellow and blue arrows), the GNR chirality can be unambiguously determined. The GNR in Fig. S2 is an (8, 1) GNR, corresponding to chiral angle of $\theta = 5.8°$.

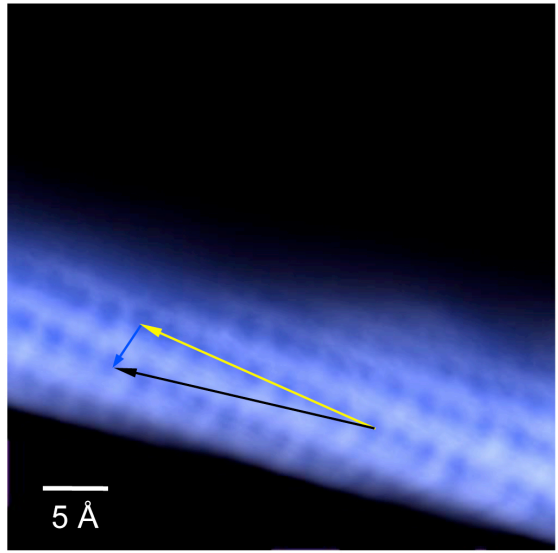

**Figure S2 | Determination of chirality of GNRs.** STM image of an (8, 1) GNR ($V_s$ = 0.3 V, I = 60 pA). The lower part is the Au surface, and the upper part is the GNR. The black line is parallel to the edge orientation, connecting two neighboring equivalent sites. The yellow and blue arrows are along the zigzag directions of graphene lattices. The

edge periodicity is 8α along the yellow line (α is the lattice constant of graphene), and 1α along the blue line.

## 5. *dI/dV* spectra of a (5, 2) GNR

In the manuscript we show a series of spectra for an (8,1) GNR. Here, in Fig. S3, we show an additional series of *dI/dV* spectra for a (5, 2) GNR with width of 15.6 ± 0.1 nm, measured at different points with lateral interval of 2.2 Å along a line perpendicular to the GNR edge (T = 7K). The energy gap here is 27.6 ± 1.0 meV and dashed lines show that the gap is unchanging with position, even though edge state amplitude decreases as one moves further into the interior of the GNR.

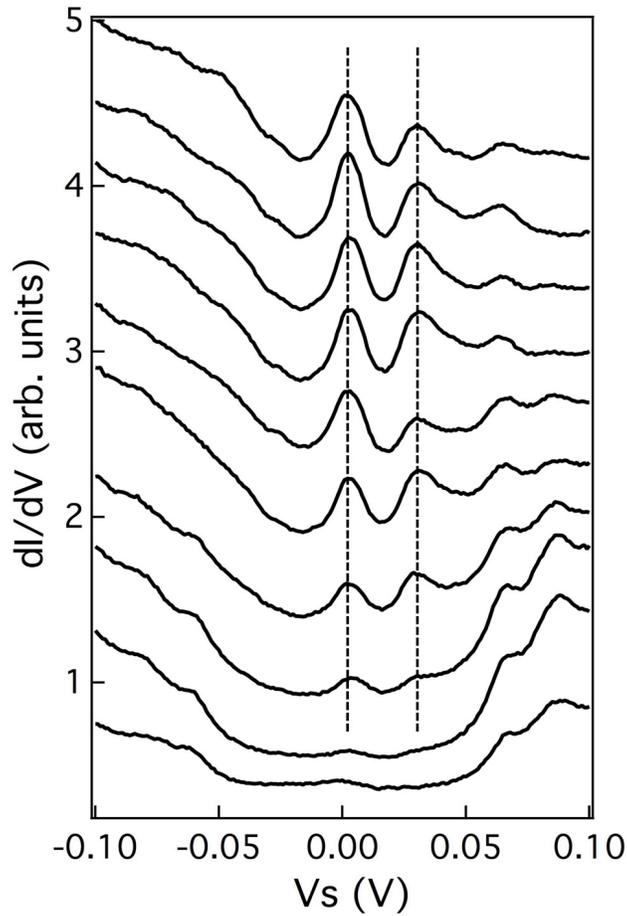

**Figure S3 | d*I*/d*V* spectra of a (5, 2) GNR**. The initial tunnelling parameters are $V_s$ = 0.15 V and *I* = 50 pA. The wiggle voltage is $V_{rms}$ = 2 mV. Dashed lines are a guide to the eye. Top curve is obtained a distance of ~ 1.0 Å from the terminal edge of the GNR (other spectra obtained at lateral intervals of 2.2 Å).